\documentclass{article}
\usepackage{frascatiphys}
\usepackage{graphicx}
\begin{document}
%
\title{ 
INDIRECT DARK MATTER SEARCH WITH LARGE NEUTRINO TELESCOPES
}
\author{
Paolo Fermani \\
on behalf of the ANTARES collaboration \\
{\em "Sapienza" Universit\`a di Roma, P.le Aldo Moro 5, 00185, Roma, Italy}\\
{\em INFN Roma, P.le Aldo Moro 2, 00185, Roma, Italy}\\
}
\maketitle
\baselineskip=11.6pt
\begin{abstract}
Dark matter is one of the main goals of neutrino astronomy. At present, there are two big neutrino telescopes based on the Cherenkov technique in ice and water: IceCube at the South Pole and ANTARES in the northern hemisphere.
Both telescopes are performing an indirect search for Dark Matter by looking for a statistical excess of neutrinos coming from astrophysical massive objects. This excess could be an evidence of the possible annihilation of dark matter particles in the centre of these objects. In one of the most popular scenarios the Dark Matter is composed of WIMP particles. The analysis and results of the ANTARES neutrino telescope for the indirect detection of Dark Matter fluxes from the Sun are here presented, as well as the latest IceCube published sensitivity results, for different Dark Matter models.
\end{abstract}
\baselineskip=14pt
\section{Introduction: Dark Matter and the WIMP scenario}
In recent years the abundance of cosmological data, such those provided by the WMAP satellite observations\cite{WMAP} or by the studies of Ia supernovae\cite{snIa}, confirmed that only the 26\% ($\Omega_m$ = 0.26) of the energy balance of the Universe is under the form of matter. Moreover, in this percentage, only the 19\% is composed by baryonic matter ($\Omega_b$ = 0.044). This means that the 81\% of matter in our Universe has a non baryonic nature ($\Omega_{cdm}$ = 0.21). This component, called Dark Matter (DM) since not luminous, is necessary to explain a lot of phenomena: among others the rotational curves in spiral galaxies\cite{rotational} and the Bullet cluster merging\cite{bullet}.  

Candidates for DM must be massive, neutral, stable on cosmological time scales and only weakly and gravitationally interacting. Since neutrinos have relativistic velocities they can not be interpreted as possible DM candidates because they were not able to form the current structures of the Universe. Thus, no Standard Model particles share these properties. One of the most popular and tested scenario is that of Dark Matter composed of the Weakly Interacting Massive Particles (WIMPs). The previous listed characteristics can be reproduced in several models\cite{bertone}. In this paper we take into account the two most accredited models: Supersymmetry (SUSY) and Universal Extra Dimensions (UED), in particular their constrained versions: CMSSM and mUED, where the WIMP particles (lightest neutralino and lightest Kaluza-Klein particle) are stable due to the conservation of R-parity and of KK-parity respectively.

There are two ways to experimentally detect WIMPs of our galactic halo\cite{comparison}.
The first is the direct search, which aim is to detect the recoil energy of the nuclei; the second is the indirect search, based on the detection of the products of WIMP annihilations in massive celestial bodies (stars, planets, galaxies). 
WIMPs loose energy through elastic scattering on nuclei and can be gravitationally captured in massive objects like the Sun, where they reach the inner core and there they can self-annihilate (being Majorana particles) producing some Standard Model particles that eventually decay producing neutrinos\footnote{The interaction producing directly neutrinos (of the same energy of the WIMP mass) is suppressed for non relativistic particles in the CMSSM case and permitted in the mUED one.} that can be observed at Earth with large neutrino telescopes.

\section{Large neutrino telescopes}
Several neutrino telescope have been built in recent years. This kind of telescopes are based on the detection of the Cherenkov light induced by the propagation of relativistic muons generated by neutrino interactions in transparent media like water or ice. Given the small value of neutrino cross-section and the decrease of the flux with the increasing energy, detectors of big mass are necessary. Here we focus on the IceCube and ANTARES experiments. 

Since the Earth acts as a shield against all the particles except neutrinos, a neutrino telescope mainly uses the detection of up-going muons as a signature of a muon neutrino ($\nu_{\mu}$) interaction in the Earth below the detector. The muon, travelling in water or ice, induces Cherenkov light that can be detected by the optical modules. The big range of distance covered by muons, permit to observe also interactions that happened hundreds meters far from the detector. Neutrinos of different flavours can also be detected, but with less efficiency and angular precision because the travelled distances of the respective leptons are short. IceCube can detect muons with a minimum energy of 10-20 GeV with the DeepCore extension, while ANTARES is able to detect muons with a minimum energy of 20 GeV and also the neutrino direction with an accuracy of roughly 0.3$^{\circ}$ for energies beyond 10 TeV.

\subsection{The IceCube experiment}
IceCube is the largest neutrino observatory in the world\cite{icecube}. It is located at the South Pole immersed in the ice. It consists of 4800 optical sensors or Digital Optical Modules (DOMs) installed on 80 strings between 1450 m and 2450 m below the surface: 1 km of instrumented lines. The in-ice array is complemented by a surface array, IceTop, composed of 160 ice-tanks at the top of the strings. Each tank contains two DOMs. There is also a more dense region of further 6 strings called DeepCore implemented in order to have an improved light collection and a lower energy threshold. The detector construction has been completed in 2011. IceCube covers an area of roughly 1 km$^2$ and a volume of 1 km$^3$. Each DOM consists of a 25 cm photomultiplier tube (PMT) in a glass sphere, equipped also with the electronics to perform the digitization and transmission of the signal to the surface such as to operate as a complete and autonomous data acquisition system. For each string there are 60 DOMs with a spacing of 17 m between them. The strings of modules are deployed in ice into holes drilled with hot water. The absolute positioning of the DOMs is measured with the deeper pressure sensor and by the means of laser range at the moment of the deployment. Depths of individual DOMs are determined to an accuracy of 50 cm. The absolute time accuracy on the time of arrival of the signals is guaranteed by a 20 MHz local clock.

\subsection{The ANTARES detector}
ANTARES is the first and largest submarine neutrino telescope in the Northern hemisphere\cite{antares}. It was completed in 2008 and located (42$^{\circ}$ 48' N, 6$^{\circ}$ 10' E) in the Mediterranean Sea at roughly 2475 m depth, 42 km offshore of the cost of Toulon (France). The detector consists of a three-dimensional array of 885 10" PMTs disposed in 12 vertical strings. These strings are spread over an area of about 0.1 km$^2$. The basic unit of the detector is the Optical Module (OM), containing one PMT and the associated electronics\cite{frontend}, housed in a pressure resistant glass sphere with a mu-metal cage to minimize the effect of the Earth's magnetic field. The OMs are grouped together in 25 storeys (of three OMs) for each string interconnected via an electro-mechanical cable with the exception of one string which has 20 storeys since the last five are devoted to acoustic measures. The OMs are arranged with the axis of the PMT tubes 45$^{\circ}$ below the horizontal plane in order to increase the efficiency to detect up-going events. The height of the instrumented strings is 300 m. The distance between two consecutive storeys is 14.5 m. The horizontal distance between two adjacent strings is 60-75 m. There is also one instrumented line for sea environmental measurements. The top of the string consists of a buoy and they are anchored on the sea bed. The absolute position of the detector components as a function of time is obtained through an acoustic triangulation system combined with an orientation system that permit to determine the inclination and orientation of the single storeys\cite{positioning}. The absolute UTC time accuracy is guaranteed by a GPS system and by the 25 MHz clock of the detector.

\section{ANTARES search for Dark Matter towards the Sun}
Here we present the analysis performed in the ANTARES collaboration to search for a signal towards the Sun with the data taken by the experiment in the period 2007-2008. The Sun is a very interesting source for Dark Matter search. In fact, the possible signal identification would have a crucial importance: we do not expect such kind of signal arising from the Sun since the solar neutrinos have a lower energy and the neutrinos produced in the Sun's corona by cosmic rays are negligible. References for other analyses with the same data sample performed in the collaboration are \cite{pointsources, monopoles, oscillation}.

\subsection{Data and simulations of signal and background}
In the large part of the year 2007 ANTARES was in a 5-lines configuration for a total of 185.5 days of active detector; while in the year 2008 the detector configuration varied: 10,9,12 (since May) lines (for maintenance and repair operations) for a total of 189.8 days of active detector. 

\begin{figure}[!htb]
    \begin{center}
        {\includegraphics[scale=0.5]{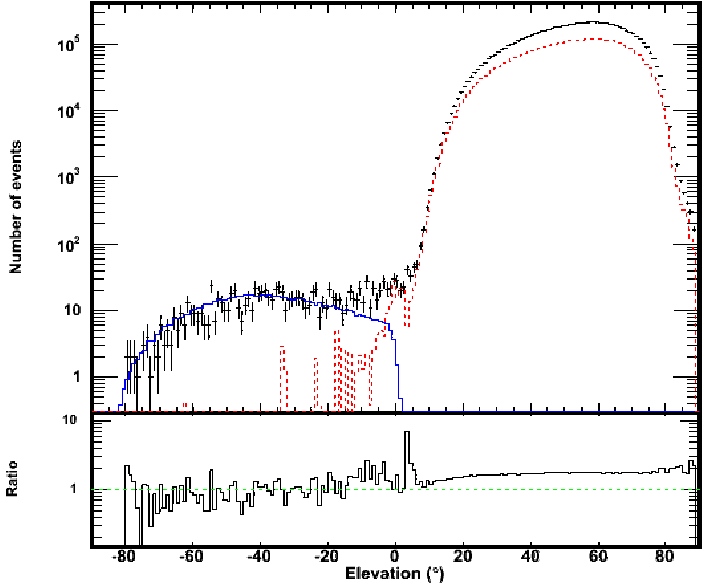}}
        \caption{\it Comparison between data and simulation for the elevation coordinate of the events. A $Q<$~1.4 is applied. The red dashed line represents the simulated atmospheric muons; the blue pointed line shows the simulated atmospheric up-going neutrinos; the black crosses represent scrambled data. The ratio of data over the simulation is shown below the main plot.}
	\label{fig:dataMC}
    \end{center}
\end{figure}

The first task in the ANTARES analyses is the discrimination between signal and background events. The main two background sources are the large flux of atmospheric muons and the flux atmospheric neutrinos both produced in the interaction of cosmic rays in the Earth's atmosphere. In order to reduce the first kind of background, the detector is installed at large depth. Moreover, only upgoing events are accepted. Still, a small fraction (but large in absolute number) of atmospheric muons are reconstructed as upgoing. By imposing strict quality cuts in the tracks, they can be further removed. The second background is irreducible since neutrinos can pass trough the Earth detected as up-going events; but we can consider that atmospheric neutrinos are isotropically distributed all over the sky while the signal neutrinos are expected to peak in the Sun direction only, so we are looking for an excess of events over an expected background. In figure \ref{fig:dataMC} an example of data-Monte Carlo comparison where the two kinds of backgrounds are involved it is shown. We have to notice that for the background estimation we used the scrambled data\footnote{Scrambled data are obtained randomising the UTC time of the events in the considered data taking period} to reduce the effects of the possible systematic uncertainties. 

To reconstruct both data and Monte Carlo events a fast and reliable algorithm, called BBFit, has been developed in the ANTARES collaboration\cite{bbfit}. This algorithm is based on the multiple coincidences of the Cherenkov photons arriving on the OMs of the apparatus (hits). Then, the minimization of a $\chi^2$ function, containing the difference between expected and measured hits, permits to reconstruct the track of the events with a given quality $Q$.

The simulation of signal events from Dark Matter annihilation in the Sun is computed with the WimpSim package\cite{wimpsim}, with which we can evaluate the differential neutrino spectra. Several annihilation channels are available for different WIMP masses in order to reproduce any possible Dark Matter model. The neutrino interactions in the Sun medium, the regeneration of $\tau$ leptons together with the standard neutrino oscillation scenario has been taken into account. We assume also that capture and annihilation rates are in equilibrium in the Sun. In the CMSSM the main annihilation channels are $W^+W^-$, $\tau^+\tau^-$ and $b\bar{b}$; for mUED are $c\bar{c}$, $\tau^+\tau^-$, $b\bar{b}$, $t\bar{t}$ and $\nu\bar{\nu}$. Another thing to note is that, although the Sun have a size of roughly half degree, since the annihilation reactions happens in its core, we can consider it as a point source.
\begin{figure}[!htb]
\begin{center}
  {\includegraphics[scale=0.5]{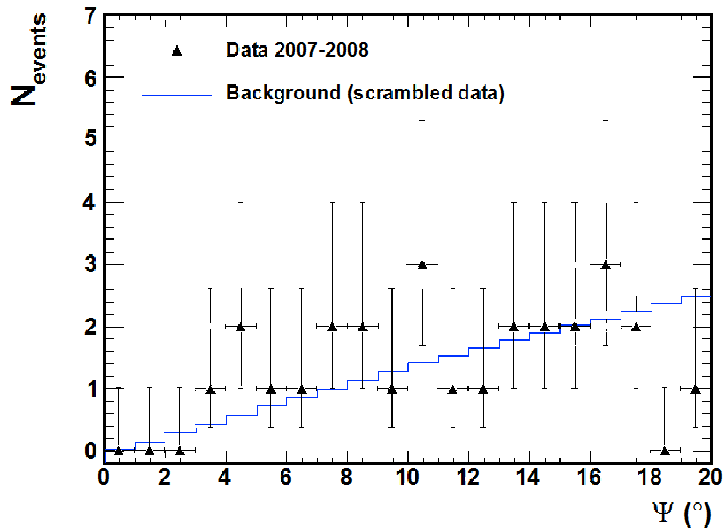}}
  \caption{\it Distribution of the spatial angle $\Psi\in$[0$^{\circ}$,20$^{\circ}$] of the event tracks with respect to the Sun's direction for the expected background computed from the time-scrambled data (solid blue line) compared to the data after the basic selection criteria (black triangles). A 1$\sigma$ Poisson uncertainty is shown for each data point (black crosses).}     
  \label{fig:unblind}
\end{center}
\end{figure}
\subsection{The optimization of the cuts}
In the analysis we performed a binned search: it means to count the number of events (signal and background) within an angular bin of given aperture $\Psi$ centred in the Sun position (see figure \ref{fig:unblind}). It is important to note that we followed also a blind\footnote{In this way the cuts are selected before to look at the source of interest, avoiding a possible bias.} procedure to choose the cuts to apply to our data. The optimization of these cuts has been done using the Model Rejection Factor (MRF) method\cite{mrf}. Two parameters were considered for the optimization: the quality of the event reconstruction $Q$ and the half-cone angle aperture around the Sun $\Psi$. The MRF give us the optimized set of these two parameters to obtain the better average upper limit (at 90\% C.L.) on the flux of neutrinos\footnote{With neutrinos we means the sum of neutrinos and anti-neutrinos.} arising from Dark Matter annihilation in the Sun. This average flux upper limit can be expressed by:
\begin{equation}
\bar{\phi}_{\nu}^{90\%}=\frac{\bar{\mu}^{90\%}}{A_{eff}(M_{WIMP})\times T_{eff}}
\label{eq:mu90}
\end{equation}
where $\bar{\mu}^{90\%}$ is the average upper limit in the event number, derived from the Feldman and Cousins calculations\cite{fc}, A$_{eff}$(M$_{WIMP}$) is the effective area and T$_{eff}$ is the active detector data taking period. This evaluation has been done for each mass and channel of the two models chosen.

\subsection{Results and conclusions}
\begin{figure}[!htb]
\begin{center}
  {\includegraphics[scale=0.6]{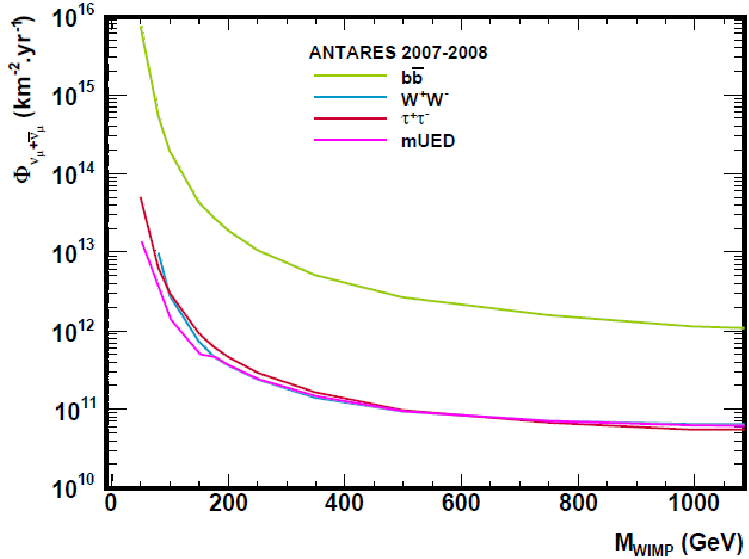}}
  \caption{\it 90\% CL upper limit on the neutrino plus anti-neutrino flux as a function of the WIMP mass in the range MWIMP$\in$[10 GeV;1 TeV] for the three channels $b\bar{b}$, $W^+W^−$, $\tau^+\tau^−$ (CMSSM) and a combination of channels (mUED).}     
  \label{fig:sensitivity}
\end{center}
\end{figure}

Using equation \ref{eq:mu90} it is possible to evaluate the sensitivity for all the Dark Matter models considered. In figure \ref{fig:sensitivity} the average upper limits in the neutrino flux as a function of the WIMP mass are shown. 
Because of the high dependence of branching ratios over the CMSSM parameter space the different channels are separate; this is not in the mUED case so we can combine the channels conserving a reliable representation. The best limits in CMSSM arise from the $W^+W^-$ and $\tau^+\tau^-$ channels since they have a hard energy spectrum (more neutrino events). For mUED case the channel that most contribute is the $\tau^+\tau^-$ so the total sensitivity is close to the one of that channel of the CMSSM case. We can also pass from these limits on neutrino fluxes to the limits on the spin-dependent cross-section of the WIMPs with protons $\sigma_{H,SD}$. The differential neutrino flux is:
\begin{equation}
\frac{d\phi_{\nu}}{dE_{\nu}}=\frac{\Gamma}{4\pi d^2}\sum_f B_f\left( \frac{dN_{\nu}}{dE_{\nu}}\right) _f
\label{eq:flux}
\end{equation} 
where $d$ is the distance between the Sun and the Earth, $\left( dN_{\nu}/dE_{\nu}\right) _f$ is the differential number of neutrino events for each channel, $B_f$ the relative branching ratios and $\Gamma\simeq C_{\otimes}/2$ is the annihilation rate as related to the capture rate $C_{\otimes}$ assuming the equilibrium of the two rates inside the Sun\footnote{The capture rate expression, assuming a Navarro Frank and White (NFW) profile for the Dark Matter density and a Maxwell-Boltzmann velocity distribution, is: $C_{\otimes}\simeq3.35\times10^{18}s^{-1}\times\left(\frac{\rho_{local}}{0.3GeVcm^{-3}}\right)\times\left(\frac{270kms^{-1}}{v_{local}}\right)\times\left(\frac{\sigma_{H,SD}}{10^{-6}pb}\right)\times\left(\frac{TeV}{M_{WIMP}}\right)^2$ .}. 
\begin{figure}[!htpb]
    \begin{center}
        {\includegraphics[scale=0.6]{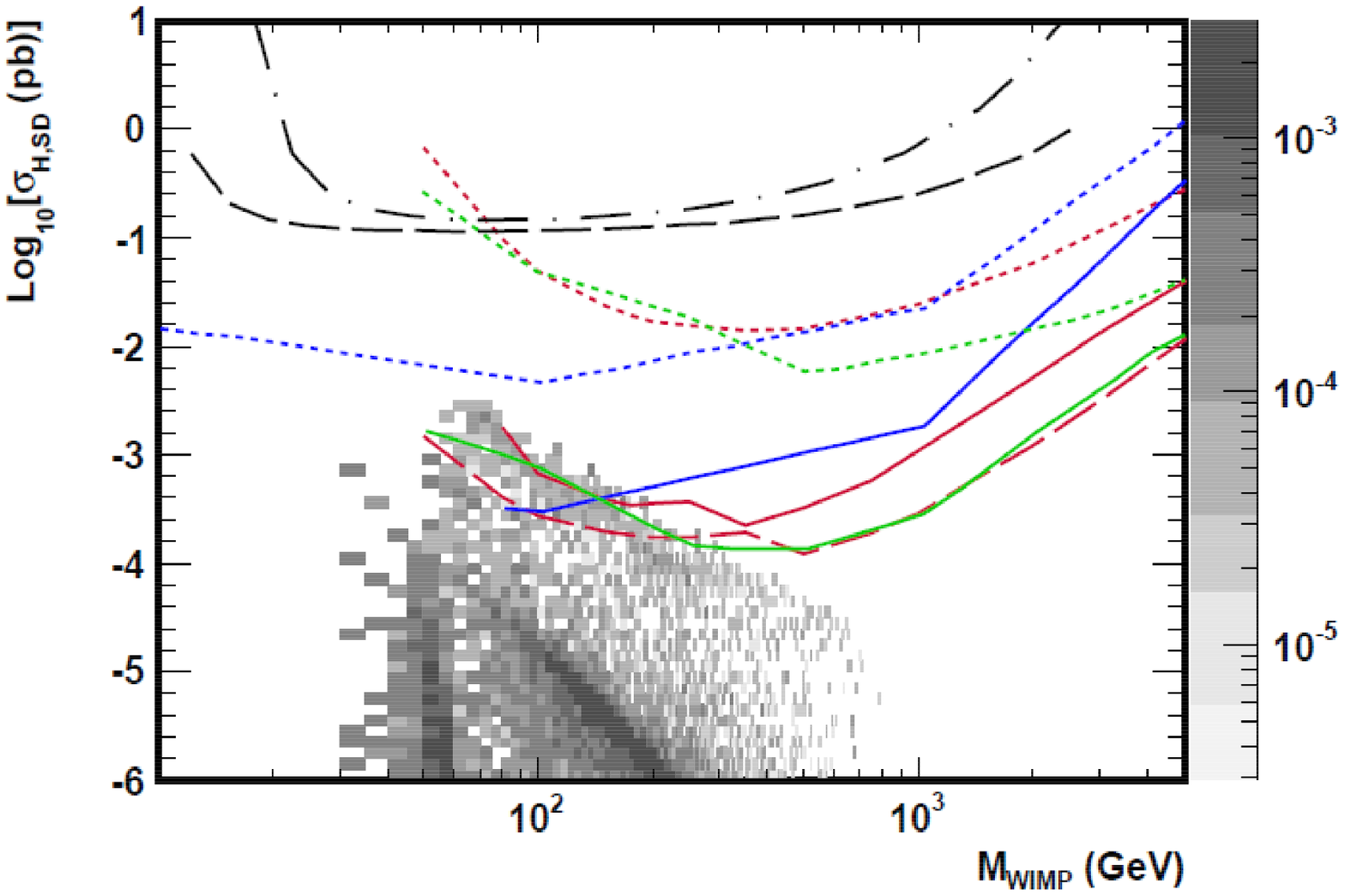}}\hspace{0.001cm} 
        {\includegraphics[scale=0.6]{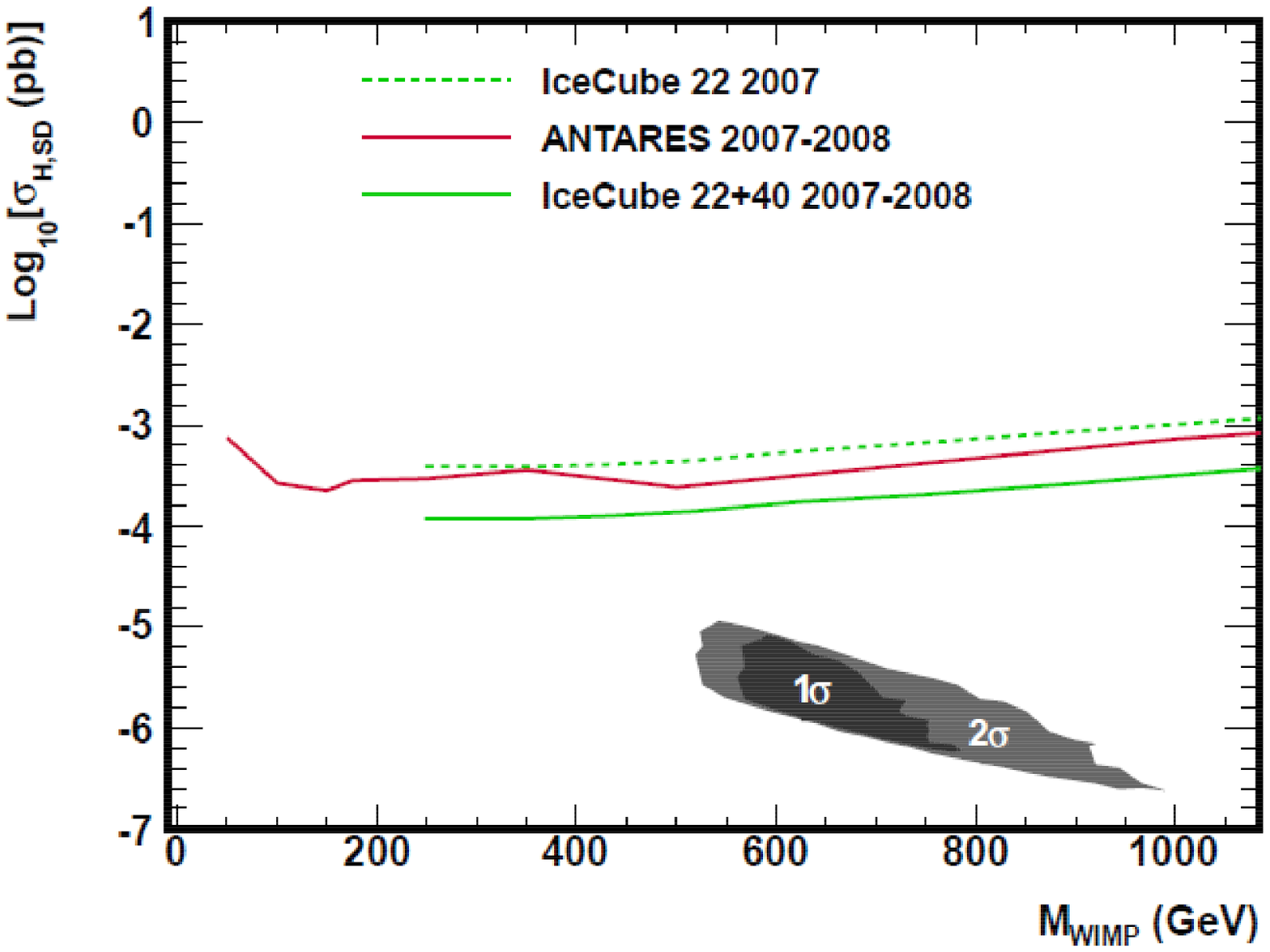}}
        \caption{\it up) 90\% CL upper limits on the spin-dependent WIMP-proton cross-section as a function of the WIMP mass in the range M$_{WIMP}\in$[10 GeV;5 TeV], for the three channels: $b\bar{b}$ (dotted line), $W^+W^−$ (solid line), $\tau^+\tau^−$ (dashed line), for ANTARES (red line) compared to the results of other indirect search experiments: SuperKamiokande 1996−2008 (blue line) and IceCube-40 plus AMANDA 2001-2008 (green line) and the results of direct search experiments: KIMS 2007 (black dot-dashed line) and COUPP 2011 (black dashed line); down) The same plot for a combination of channels in the mUED model (red line). Results from IceCube-22 2007 (green dotted line) and IceCube-22+40 2007-2008 (green solid line) are shown for comparison.}
	\label{fig:SD}
    \end{center}
\end{figure}

In figure \ref{fig:SD} the limits on the $\sigma_{H,SD}$ values (obtained with a scan of the SuperBayes package\cite{superbayes}) for CMSSM and mUED models can be seen. In the spin-dependent case both ANTARES and IceCube limits are competitive compared with the direct search experiments (this is not in the case for limits on the spin-independent cross-section). 
\begin{figure}[!htb]
    \begin{center}
        {\includegraphics[scale=0.4]{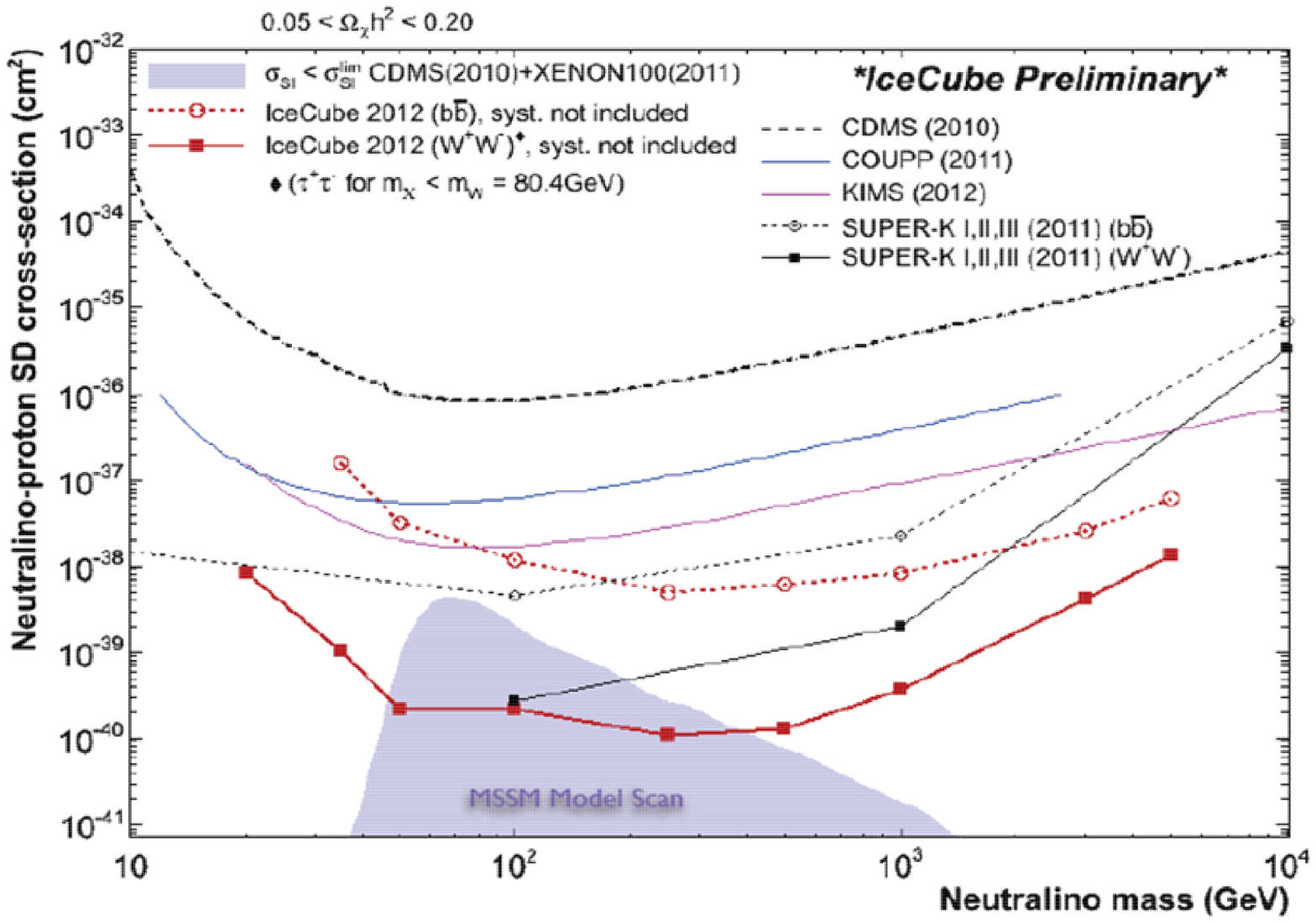}}
        \caption{\it 90\% CL upper limits on spin-dependent WIMP-proton cross-section as a function of the WIMP mass for annihilation channels $W^+W^-$ (red solid line) $\tau^+\tau^-$ (black square and line) and $b\bar{b}$ (red dotted line) over a range of WIMP masses. Systematic uncertainties are included. The shaded region represents an allowed MSSM parameter space taking into account recent accelerator, cosmological and direct DM search constraints. Results from Super-K, COUPP, KIMS, CDMS, XENON-100 are shown for comparison\cite{danninger}.}
	\label{fig:ice}
    \end{center}
\end{figure}

\section{Summary of the IceCube Sun analysis}
The search for Dark Matter in the IceCube collaboration has given results with the 79-string detector\cite{ice79} and, for the first time, with the DeepCore expansion in the period between June 2010 and May 2011\footnote{The data acquisition is divided in two periods: austral summer and austral winter, when the Sun is above and below the horizon respectively}. The total active detector lifetime was of 317 days, with more than $60\times10^9$ recorded events and roughly 25000 signal-like events in the final state. With DeepCore the analysis has reached neutrino energies of 10-20 GeV. To simulate the signal the packages DarkSusy\cite{darksusy} and WimpSim\cite{wimpsim} were used. The search was binned with respect to the Sun position. Comparing signal simulation and data some cuts have been placed to reduce the contamination of atmospheric muon events (requiring a good quality reconstruction and a strong containment detection). To optimize the applied cuts a likelihood analysis was done (with the Feldman and Cousins technique\cite{fc}). In figure \ref{fig:ice} the results of this analysis, as limits on the WIMP-proton spin-dependent cross-section, compared to other experimental limits (for both direct and indirect searches) are shown. 
For a more complete review on the search for Dark Matter in the IceCube collaboration see the reference \cite{iceref}.
%

%
\end{document}